# It should be Einstein-Laub 'Equations' inside Matter

Mahdy Rahman Chowdhury Mahdy

Department of ECE, National University of Singapore

*Abstract*— In a recent article [1], Mansuripur has claimed that inside the matter conventional Lorentz Force law should be abandoned in favour of a more general expression of the electromagnetic force density such as the one discovered by A. Einstein and J. Laub in 1908. The main focus of Mansuripur's claim is based on special theory of relativity. According to [1], Lorentz force law is incompatible with special theory of relativity. In this article, we have focused in favour of Einestien-Laub equations (Force law; associated stress tensor, momentum density, Poynting vector etc) from quite different point of views. Especially we have tried to include previously/recently reported experimental observations, Abraham-Minkowski –Nelson controversy, Quantum Electrodynamics and most importantly the significance of associated stress tensors to judge the problem from a broader and engineering point of view. At the end of the day, considering all the issues, we have found that only Einestien-Laub 'equations' can predict the actual total force inside the matter and probably consistent with all other laws of Electrodynamics.

According to [1], the conventional Lorentz force law should be rejected due to incompatibility with special relativity [2] and due to the 'problem'- 'hidden momentum' (Shockley version [3],[4]). Several comments have been made against Mansuripur's claim ([5], [6] & references there in). Recently Mansuripur has responded to critics in his one article [7]. Actually, mansuripur's claims are too much theoretical and somehow they have missed some important issues like previously/recently reported experimental observations, Abraham-Minkowski–Nelson controversy, Quantum Electrodynamics (QED) and most importantly the significance of associated stress tensors. In this article we have tried to support Einestien-Laub force inside the matter from completely different point of views. The most important issue is that- if a theory/ law is good enough to handle the problem from one point of view, it should be successful with all associated laws and theories. Probably from four vector point of view, it is possible to draw a half-successful conclusion against 'Mansuripur paradox' (relativity problem) [7]. But when we consider the associated stress tensor of conventional Lorentz force law, we face some serious problems to calculate the total force (see our recent article [8]). Besides, when we consider a much complicated case-'bounded magneto dielectric particle embedded in unbounded magneto dielectric background', conventional Lorentz force law seriously fails to handle such complicated cases [8]. Only by modifying the Einestien-Laub equations (force law, associated stress tensor, momentum density, Torque equation etc), it may be possible to handle such complicated cases [8]. However, in this article, we have noted our views very shortly that support Einestien-Laub 'equations' but contradict with conventional Lorentz force and its associated equations:

(1) Maximum comments against Mansuripur Paradox [1] have missed one important point: If conventional Lorentz force law is correct one inside the matter, it should be consistent with momentum conservation law and total force calculation based on appropriate stress tensor and momentum density. Unfortunately Nelson stress tensor [9] which is the stress tensor of conventional Lorentz force is not a good choice to calculate the total force [8],[10] & see Figure-1.

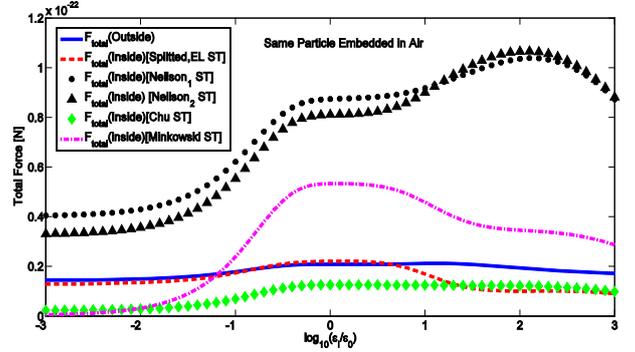

Figure-1: From [8]. Background is air. $\epsilon_s = 7 + I*\epsilon_I$ & $\mu_s = 3$ of the magneto dielectric spherical mie particle (detail data and calculation available in [8]). Total outside force has been calculated by using conventional Maxwell stress tensor. Total inside forces has been calculated by using only internal fields following different stress tensors (Einestien-Laub ST, Chu ST , Minkowski ST , Nelson1 ST (equation(3.18) of [9] ) , Nelson2 ST(equation (2.20) of [9]); also see table1 of [21]). Only the total internal force of EL ST is matching with outside total force up to loss level 10 [in log scale '1']. However, Stress tensor of conventional Lorentz force (Nelson ST s) stands far away from the actual total force. r=0.999a for inside calculation.

On the other hand, although Mansuripur has always claimed to avoid the 'hidden momentum' (Shockley version [3], [4]), we have shown 'hidden momentum' as a part of the total force (but in kinetic force part; see equation (8) and (9)). We argue that EL force law handles such hidden momentum in a different way. For this reason, it is not necessary to count that hidden momentum in the EL force law (see later). Probably Mansuripur has missed this important point. In several comments against [1], it has been proposed that- Amperian current loop should naturally rise hidden momentum due to relativistic effect [5],[6],[11] and the total linear relativistic momentum has been conserved by its addition[11]. By adding the contribution of that hidden momentum with conventional Lorentz force manually, it is possible to calculate the total force [12]. But In [13], it has been shown addition of that extra hidden momentum part with conventional Lorentz force ultimately gives nothing but the EL force which supports Abraham Momentum Density (AMD)with EL stress tensor (not Nelson one):

$$F_{CL}(r,t) = \rho_{total} E + J_{total} \times B \qquad (1)$$

$$\rho_{total} = \rho_{free} - \nabla \cdot P = \text{Free charges + bound charges} \qquad (2)$$

$$J_{total} = J_{free} + \partial P/\partial t + \left(1/\mu_o\right) \cdot (\nabla \times M)$$

= free + bound currents  (3)

Considering complete differentials; after algebraic calculations-

$$F_{CL}(r,t) = [\rho_{free} E + J_{free} \times \mu_o H + (P \cdot \nabla)E + \left(\partial P/\partial t\right) \times \mu_o H$$

$$+ (M \cdot \nabla)H - \left(\partial M/\partial t\right) \times \epsilon_o E - \epsilon_o \frac{\partial}{\partial t}(E \times M)$$

= (EL force – Hidden momentum contribution)  (4)

However, we state hidden momentum as a 'problem' because the calculation of that time varying hidden momentum may not give correct force, we are not interested to add it manually and it can be explained from quite different point of view using QED (see equation (8) and (9)). Probably neither Lorentz force law nor EL force law explains its actual origin (see equation (8)). But EL force handles it successfully in an alternative way. Although EL force may not be the law of nature [14, 15] (even conventional Lorentz force may not be [14]), EL Stress tensor/force can calculate the total force quite successfully just by using inside electromagnetic field [8]. We have also shown that such force calculation using any other force law/stress tensor (including Nelson stress tensor) stands far away from correct result [8] (see Figure-1). Besides, the definition of Poynting vector is also not free from controversy [15] for Nelson one. Nelson momentum density has not been observed in any practical experiment, although some recent articles predict in favour of it [16, 17] considering Abraham momentum density as its alternate. However, in [16, 17]; $\mu=1$; so the argument needs further verification. If somehow the force will be found like this- $(\epsilon_r-1)$ Instead of $(\epsilon_r-\mu_r^{-1})$, only then the presence of Nelson momentum density will be proved. But it would be probably the only example in favor of such density. There are several other experiments which support intrinsic Abraham force and Abraham momentum density. We have discussed the issue later in this article. According to our prediction; the force found in [17], is nothing but intrinsic Abraham force supporting AMD. Moreover, the 1973 Ashkin and Dziedzic experiment [18] shows poor agreement with conventional Lorentz law [19] but it supports the theoretical analyses based on EL formula [20].

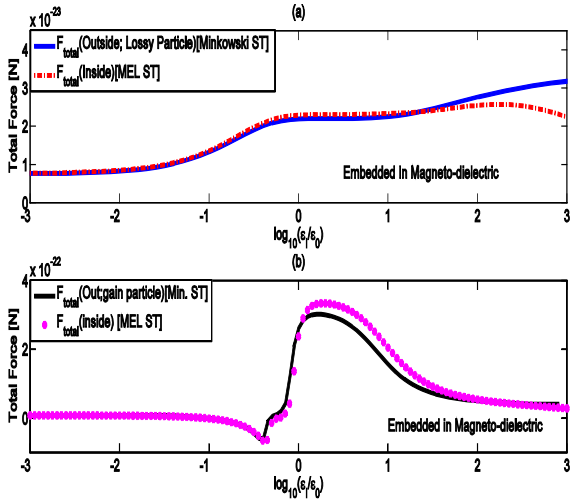

Figure-2: From [8].Magneto dielectric Background (detail data and calculations are available in [8]) for both the spherical mie particles (lossy and gain particle). Radius of each spherical mie particle is 1000 nm. The total outside force has been calculated by applying Minkowski stress tensor supporting canonical Momenta [8],[21,22],[24,25],[27],[37]. The total inside forces has been calculated applying Modified Einstein Laub stress tensor (proposed in [8]) by using internal fields only. (a)Lossy Particle: $\epsilon_s = 15+ I*\epsilon_I$; $\mu_s$=3. MEL shows only maximum 2% error in some regions and handles all loss values successfully (extremely efficient). (b)Gain particle: $\epsilon_s = 15- I*\epsilon_I$ ; $\mu_s$=3. MEL stress tensor works efficiently and validates the idea of optical pulling force. Set r=0.999a for inside calculation.

(2) In [6], another solution has been proposed against EL force considering Gilbert dipole and writing the asymmetric force equation-

$$f_{GD} = \rho^* B - \frac{1}{c^2} J^* \times E \qquad (5)$$

Its associated stress tensor will face the same problems like Nelson stress tensor to calculate the total force. Its associated Poynting vector, electric displacement vector, D, and momentum density also face similar problems stated in [15]. But Einstein-Laub formulas always provide a 'recipe' for calculating the correct force associated with kinetic momenta and Abraham Momentum density (AMD) by supporting symmetric splitted system(Maxwell-Chu like formulations[21],[22]). The constitutive relations, Poynting vector and associated torque give consistent result with the symmetric EL force [1]:

$$f_{EL} = (P\cdot\nabla)E + (M\cdot\nabla)H + J_e \times \mu_o H - J_m \times \epsilon_o \overline{E} \qquad (6)$$

$$J_e = \partial P/\partial t \; , \; J_m = \frac{\partial M}{\partial t} \qquad (7)$$

(3) The main problem associated with the forces supporting kinetic momena and AMD is- to consider the existence of magnetic monopole (Maxwell-Chu equations [21],[22] to describe a symmetric splitted system [8])which has not been found so far experimentally [23]. But we can give a different look to this 'magnetic monopole' issue. The fact that the electron orbits inside atoms and molecules are stable is a quantum-mechanical phenomenon. Neither Maxwell's equations nor the Lorentz law of force (1) or (5) (and nor, for that matter, the Einstein-Laub force/torque equations) can account for the stability of the electron orbit. The fact that electrons, protons and neutrons have a magnetic moment associated with their spin angular momentum is also a relativistic quantum effect that has no explanation within classical physics. Still such force laws (Abraham or Chu or EL; see table1 of [21]) give us a correct result just using inside electromagnetic field because the total probabilistic momentum conservation is supported by such formulations (section-3 of [24]) and they manage the quantum mechanical Aharonov casher (or any similar effect equivalent to classical 'hidden momentum'; see the references in [12]) and Rontzen effect classically [24],[25]. It can be shown that- the physical origin of such mechanical force can be described by-

$$F_{Abraham} = \int \frac{\partial}{\partial t}(g_{Min} - g_{Abr})dV = \frac{\partial}{\partial t}\int dV\left(D\times B - \frac{E\times M}{c^2}\right) \qquad (8)$$

$$= \frac{\partial}{\partial t}\int dV\left(P\times B - \frac{M\times E}{c^2}\right) = \frac{\partial}{\partial t}\left(P_{kinetic}^{medium} - P_{canonical}^{medium}\right) \qquad (9)$$

EL force not only calculates this kinetic part (equation (8) & (9)) [17],[26,27] but also calculates the static part of the force inside the matter[26],[27].At optical frequencies the situation is less clear for Abraham force [26]. Here, the Abraham force cannot be observed directly, as it averages to zero over one cycle, and one has to resort to the observation of momentum transfer from light to matter, with the aforementioned conceptual difficulty which type of momentum to consider [17].Neither Chu nor Abraham force can calculate the static part of the force using the field inside the matter [26],[27].But EL force does not face the difficulty associated with that time varying part (hidden momentum of Abraham version [4]) and the static part of the force [26],[27]. Hence EL law (both force and stress tensor [8]) can handle the total force where the media are described not in terms of free and bound charges and currents, but as spatio-temporal distributions of charge, current, polarization and magnetization (also see equation (22) to (25) in [27]). Neither Chu nor Abraham stress tensor can calculate the total force inside the matter using only the inside electromagnetic field. But it is EL stress tensor which can calculate the total force by only using the inside electromagnetic field (see Figure-1 mainly and also Figure-2).

(4) In a recent article Rikken et al.[17] have predicted about quantum mechanical pseudo-momentum in favor of Nelson momentum density. However, the observed momentum in [17] is not the pseudo-momentum reported in [9]. This quantum mechanical pseudo-momentum [16, 17] originates from relativistic quantum vacuum effect [28]. But it predicts - 'Cashimir momentum' could become within reach of experimental observation [29]. On the other hand, Aharonov-Casher effect and Rontzen effect have already been reported from experimental observations [25],[30,31,32]. Besides, the authors have reported the case of $\mu = 1$. So, it may be better to consider the momentum density in [17] as Abraham one. Moreover, The experimental observation of the Abraham force has been reported in [33-36],[16]. So, Even if the theoretical description of [24,25],[37] is not right (for some special cases), AMD (that supports kinetic momenta) inside an object is an experimental fact. Besides, there are a lot of experimental works which also support Minkowski momentum density (associated with canonical momenta) as correct one specially for unbounded background material issue and few submerged particles [21],[27],[37] & references there in. These observations validate the idea of kinetic and canonical momenta and also equation (8) and (9). So, to enclose the sub-system the EL stress tensor should be considered as the most effective one considering all the above mentioned cases. Although, Rikken et al. [17] have argued in favor of quantum mechanical pseudo-momentum; it has been clearly stated: "the Einstein-Laub version is equivalent to the Abraham version" (from force point of view but not stress tensor view) and they have measured nothing but the intrinsic Abraham force. However, Abraham stress tensor [see Table-1 in [21]] cannot calculate the total force inside the matter because it is just symmetric version of Minkowski stress tensor to conserve the angular momentum. Mathematically Abraham stress tensor gives the same result like Minkowski stress tensor which cannot calculate the total force inside the matter. Moreover, when the static part of force starts to play vital role, no other force but EL force (and also stress tensor) can handle it along with kinetic one inside the matter (see Figure-1).

However, the main problem of EL equations is its inefficiency to describe the correct total force of submerged and embedded particles (Hakim-Higham experiment [26], the observation in [38] etc). Probably we have found a solution of this problem [8] (see Figure-2). The solution argues to modify 'only' the Einestien-Laub 'equations' not only to calculate the correct total force but also to make compact, consistent and generalized foundation for electromagnetic-kinetic systems [8]. Although Stress tensors have been considered dangerous for total force calculation [1], we can calculate total force successfully for submerged and embedded cases applying modified Einstein-Laub stress tensor proposed in [8] (see Figure-2). So, at the end we may conclude that- EL 'equations' should be considered as the most generalized equations inside the matter (particle embedded in air that supports kinetic momenta).

**Acknowledgment.** M.R.C. Mahdy thanks Dr. Qiu Cheng Wei of NUS for his support and guidance.